\documentclass[manuscript,times]{aastex62}
\usepackage{booktabs}

\renewcommand{\vec}[1]{\boldsymbol{#1}}

\shorttitle{}
\shortauthors{Jiang et al.}

\usepackage{subfigure}
\usepackage{url} 
\usepackage{amsmath,amssymb}
\usepackage{multirow}
\usepackage{graphicx} 
\usepackage{epstopdf}
\begin{document}

%
%


\title{Velocity Space Signatures of Resonant Energy Transfer between Whistler Waves and Electrons in the Earth's Magnetosheath}

%
%

\correspondingauthor{Wence Jiang}
\email{jiangwence@swl.ac.cn}

\author[0000-0002-8055-4569]{Wence Jiang}
\affiliation{State Key Laboratory of Space Weather, National Space Science Center, Chinese Academy of Sciences, Beijing, 100190, China}
\affiliation{Mullard Space Science Laboratory, University College London, Dorking RH5 6NT, UK}
\affiliation{Key Laboratory of Solar Activity and Space Weather, National Space Science Center, Chinese Academy of Sciences, Beijing,100190, China}

\author[0000-0002-0497-1096]{Daniel Verscharen}
\affiliation{Mullard Space Science Laboratory, University College London, Dorking RH5 6NT, UK}

\author[0000-0001-8529-3217]{Seong-Yeop Jeong}
\affiliation{Samsung Electronics Co. Ltd, Hwaseong, 18448, South Korea}

\author[0000-0002-4839-4614]{Hui Li}
\affiliation{State Key Laboratory of Space Weather, National Space Science Center, Chinese Academy of Sciences, Beijing, 100190, China}
\affiliation{Key Laboratory of Solar Activity and Space Weather, National Space Science Center, Chinese Academy of Sciences, Beijing,100190, China}
\affiliation{University of Chinese Academy of Sciences, Beijing, China}

\author[0000-0001-6038-1923]{Kristopher G. Klein}
\affiliation{Department of Planetary Sciences, University of Arizona, Tucson, AZ, USA}

\author{Christopher J. Owen}
\affiliation{Mullard Space Science Laboratory, University College London, Dorking RH5 6NT, UK}

\author[0000-0001-6991-9398]{Chi Wang}
\affiliation{State Key Laboratory of Space Weather, National Space Science Center, Chinese Academy of Sciences, Beijing, 100190, China}
\affiliation{Key Laboratory of Solar Activity and Space Weather, National Space Science Center, Chinese Academy of Sciences, Beijing,100190, China}
\affiliation{University of Chinese Academy of Sciences, Beijing, China}


\begin{abstract}

Wave--particle interactions play a crucial role in transferring energy between electromagnetic fields and charged particles in space and astrophysical plasmas. Despite the prevalence of different electromagnetic waves in space, there is still a lack of understanding of fundamental aspects of wave--particle interactions, particularly in terms of energy flow and velocity-space characteristics. In this study, we combine a novel quasilinear model with observations from the Magnetospheric Multiscale (MMS) mission to reveal the signatures of resonant interactions between electrons and whistler waves in magnetic holes, which are coherent structures often found in the Earth's magnetosheath. We investigate the energy transfer rates and velocity-space characteristics associated with Landau and cyclotron resonances between electrons and slightly oblique propagating whistler waves. In the case of our observed magnetic hole, the loss of electron kinetic energy primarily contributes to the growth of whistler waves through the $n=-1$ cyclotron resonance, where $n$ is the order of the resonance expansion in linear Vlasov--Maxwell theory. The excitation of whistler waves leads to a reduction of the temperature anisotropy and parallel heating of the electrons. Our study offers a new and self-consistent understanding of resonant energy transfer in turbulent plasmas.

\end{abstract}

\section{Introduction}\label{sec0}

Electromagnetic fluctuations in space and astrophysical plasmas expand across an extensive range of spatial and temporal scales \citep{Tu95,Bruno13,Alexandrova2013,Verscharen2019b,Sahraoui20}.  The interactions between charged particles and electromagnetic fluctuations play crucial roles for the energy conversion and dissipation in astrophysical plasma environments such as the solar wind, planetary magnetospheres, and the interstellar medium \citep{Marsch2006,Schekochihin2009}. Resonant wave--particle interactions include Landau-resonant and cyclotron-resonant processes. They are efficient mechanisms to convert energy between electromagnetic fields and particles, causing particle acceleration/deceleration, the kinetic evolution of the particle velocity distribution function (VDF), and turbulence dissipation \citep{Marsch1982,Gurnett1983,He2015,Xiao2015,Chen2019,Liu2022}. Non-resonant wave--particle interactions include stochastic heating and magnetic reconnection in wave fields \citep{Johnson2001,Voitenko2004,Chandran2010,Loureiro2017,Agudelo2021,Agudelo2022}. To diagnose the signature of energy transfer in spacecraft observations, insightful techniques such as the field--particle correlation technique have been developed and successfully implemented \citep{Klein2017,Howes2017,Verniero2021,Montag2023}. These methods reveal, for example, the dissipation of kinetic Alfv\'en waves through Landau damping in the Earth's magnetosheath \citep{Verniero2021,Chen2019}.

In the Earth's magnetosheath, enhanced electromagnetic fluctuations at kinetic scales such as whistler waves are sometimes localized near coherent structures like current sheets, magnetic islands, and magnetic holes \citep{Zhang1998,Tsurutani2011,Karimabadi2014,Ahmadi2018,Breuillard2018,Kitamura2020,Behar2020,Li2021}. These whistler waves are frequently observed as right-hand polarized electromagnetic waves with a small propagation angle with respect to the background magnetic field. Micro-instabilities driven by unstable butterfly or beam-like VDFs are key candidates to explain the occurrence of these waves \citep{Zhima2015,Ahmadi2018,Ren2019,Huang2020,Zhang2021}. Although the direct observation of the energy transfer via cyclotron resonance is sometimes possible through data from the Magnetospheric Multiscale (MMS) mission \citep{Kitamura2023}, the understanding of the associated velocity-space signatures and the time-dependent properties of the resonant energy transfer between electrons and whistler waves is still lacking.

In this letter, we focus on an interval previously studied by \citet{Jiang2022} and use a novel quasilinear model to numerically solve the quasilinear impact of wave--particle interactions on the temporal and energy evolution of the plasma. We discuss the quantified signatures of energy transfer between whistler waves and electrons under the action of three different wave-particle resonance mechanisms. Finally, we provide suggestions for direct in-situ observations of such resonant wave-particle interactions.

\section{Quasilinear evolution of the distribution function}

Quasilinear theory describes the collective and slow (compared to the wave frequency) response of the VDF to fluctuating electromagnetic fields in resonant wave--particle interactions \citep{Shapiro1962,Kennel1966,Rowlands1966}. In quasilinear theory, the time evolution of electron VDF via resonant interactions between electromagnetic fields and electrons (denoted by subscript $\mathrm e$) follows a diffusion in velocity space:

\begin{equation}
\frac{\partial f_\mathrm{e}}{\partial t}=  \lim\limits_\mathrm{{V \to \infty}} \sum_\mathrm{n=-\infty}^{\infty} \int \frac{\pi q_\mathrm{e}^2}{V m_\mathrm{e}^2} \hat G [k_\parallel] 
\times \left[ \frac{v_\perp^2}{\lvert v_\parallel \lvert} \delta \left( k_\parallel - \frac{\omega_\mathrm{k}-n\Omega_\mathrm{e}}{v_\parallel} \right) \lvert \psi_\mathrm{e}^{n} \rvert ^2 \hat G [k_\parallel] f_\mathrm{e} \right] d^3 \vec{k},
\label{eq1}
\end{equation}

where $f_{\mathrm e}$ is the electron VDF, 
\begin{equation}
\hat G[k_\parallel]= \left(1-\frac{k_\parallel v_\parallel}{\omega_k}  \right)\frac{1}{v_\perp} \frac{\partial}{\partial v_\perp} + \frac{k_\parallel}{\omega_k} \frac{\partial}{\partial v_\parallel},
\label{eq3}
\end{equation}
\begin{equation}
\psi_e^n= \frac{1}{\sqrt{2}} [E_k^R e^{i \phi} J_\mathrm{n+1}(\rho_e)+ E_k^L e^{-i \phi} J_\mathrm{n-1}(\rho_e)]\\
+\frac{v_\parallel}{v_\perp}E_k^z J_n(\rho_e),
\label{eq2}
\end{equation}
$q_{\mathrm e}$ is the charge of an electron, $m_{\mathrm e}$ is the mass of an electron, $k_{\parallel}$ is the wavevector component parallel to the background magnetic field $\vec B_{0}$ so that the full wavevector is decomposed as $\vec k=(k_{\perp}\cos\phi,k_{\perp}\sin\phi,k_{\parallel})$, $v_{\perp}$ is the velocity component perpendicular to $\vec B_0$, $v_{\parallel}$ is the velocity component parallel to $\vec B_0$, $\omega_k$ is the real part of the wave frequency, $n$ is an integer ($n=0$ represents the Landau resonance and $n\ne0$  represents cyclotron resonances), $J_{n}$ is the Bessel function, $\rho_e=k_\perp v_\perp /\Omega_e$, $\Omega_e=q_eB_0/(m_e c)$ is the electron cyclotron frequency, and $c$ is the speed of light. In quasilinear theory, it is assumed that $|\gamma_k| \ll |\omega_k|$, where $\gamma_k$ is the imaginary part of the wave frequency. We define the Fourier transformation of the electric field as $\vec E_k=(E_k^x,E_k^y,E_k^z)$ and its circular components as $E_k^R=(E_k^x-iE_k^y)/\sqrt{2}$ and $E_k^L=(E_k^x+iE_k^y)/\sqrt{2}$ \citep[see also][]{Verscharen2019a}.

\section{Method}\label{}

\subsection{Numerical Model for the Quasilinear Evolution}\label{}
To investigate the time-dependent nature of quasilinear diffusion, we use a numerical model to solve the time evolution of electron VDFs according to Eq.~(\ref{eq1}) \citep[]{Jeong2020}. Using a Crank-Nicolson scheme, the \citet{Jeong2020} model is a novel and generalized method to solve the time evolution of 2-dimensional VDFs under the action of a dominant resonant wave--particle resonance. 

In this model, we define approximate window functions to reflect the distribution of wave energy over $k_{\parallel}$ \citep[see][]{Jeong2020} as
\begin{equation} 
W^{n} =\frac{1}{|v_\parallel-v_\mathrm{g0}|} \exp{\left[ -\frac{k^2_\mathrm{\parallel 0}}{\sigma^2_\mathrm{\parallel 0}} \left( \frac{v_\parallel -v^{n}_\mathrm{\parallel res}}{v_\parallel-v_\mathrm{g0}} \right)^2 \right]},
\label{eq4} 
\end{equation}
where
\begin{equation} v_\mathrm{\parallel res}^n =\frac{\omega_{k0}-n \Omega_{\mathrm e}}{k_\mathrm{\parallel 0}}
\label{eq5} 
\end{equation}
is the $n$th order resonance velocity, $v_\mathrm{g0}$ is the group velocity, $k_\mathrm{\parallel0}$ is the central parallel wave number, $\omega_{k 0}$ is the central frequency, $\sigma_{\parallel 0}$ is the half width of the window function, $v_{\mathrm{Ae}}=B_0/\sqrt{4\pi n_{\mathrm e}m_{\mathrm e}}$ is the electron Alfv\'en speed, and $n_{\mathrm e}$ is the electron number density. Eq.~\ref{eq4} determines the region in $v_\parallel$ space in which the quasilinear diffusion through the $n$th order resonance with unstable whistler waves is effective. 

The width of the window function is determined by the unstable whistler-wave spectrum calculated with the Arbitrary Linear Plasma Solver \citep[ALPS;][]{Verscharen2018a,alps2023}. To determine the window functions, we implement a non-Maxwellian VDF model into ALPS to evaluate the stability of whistler waves. The VDF model is based on realistic electron VDF data from MMS1 on 2017 January 25 from 00:25:44.38 to 00:26:44.80 UT \citep[more details see][]{Jiang2022}. Figure~\ref{fig2}b shows isocontours of the initial electron VDF. ALPS predicts an unstable spectrum of whistler waves, i.e., $\gamma_k >0$ in the range $0.5 < k_\mathrm{\parallel}d_{\mathrm e} < 0.8$, where $d_\mathrm{e}$ denotes the electron inertial length. Therefore, we set $\sigma_{\parallel 0}=0.08 d_\mathrm{e}$, $k_\mathrm{\parallel0}d_e=0.64$ and $v_\mathrm{g0}= 0.07 v_\mathrm{Ae}$ in our numerical model. The unstable whistler waves have a propagation angle slightly oblique to the background magnetic field ($\sim 10^\circ$) and the real part of the frequency at maximum growth is $\omega_{k0}=0.26\Omega_{\mathrm e}$, which is in agreement with the observed wave properties \citep{Jiang2022}. We determine the magnetic-field amplitude of whistler waves as 0.0018 nT from direct MMS observations \citep{Torbert2016}. In our model, the electron VDF is discretized into a $120\times240$ grid on the ($v_\parallel,v_\perp$) plane, where $-3 v_\mathrm{Ae} \leq v_\parallel \leq 3 v_\mathrm{ Ae}$ and $0 \leq v_\perp \leq 3 v_\mathrm{Ae}$.


\subsection{Trajectories of Quasilinear Diffusion}\label{}
\begin{figure}[!htbp]
\centering
\plotone{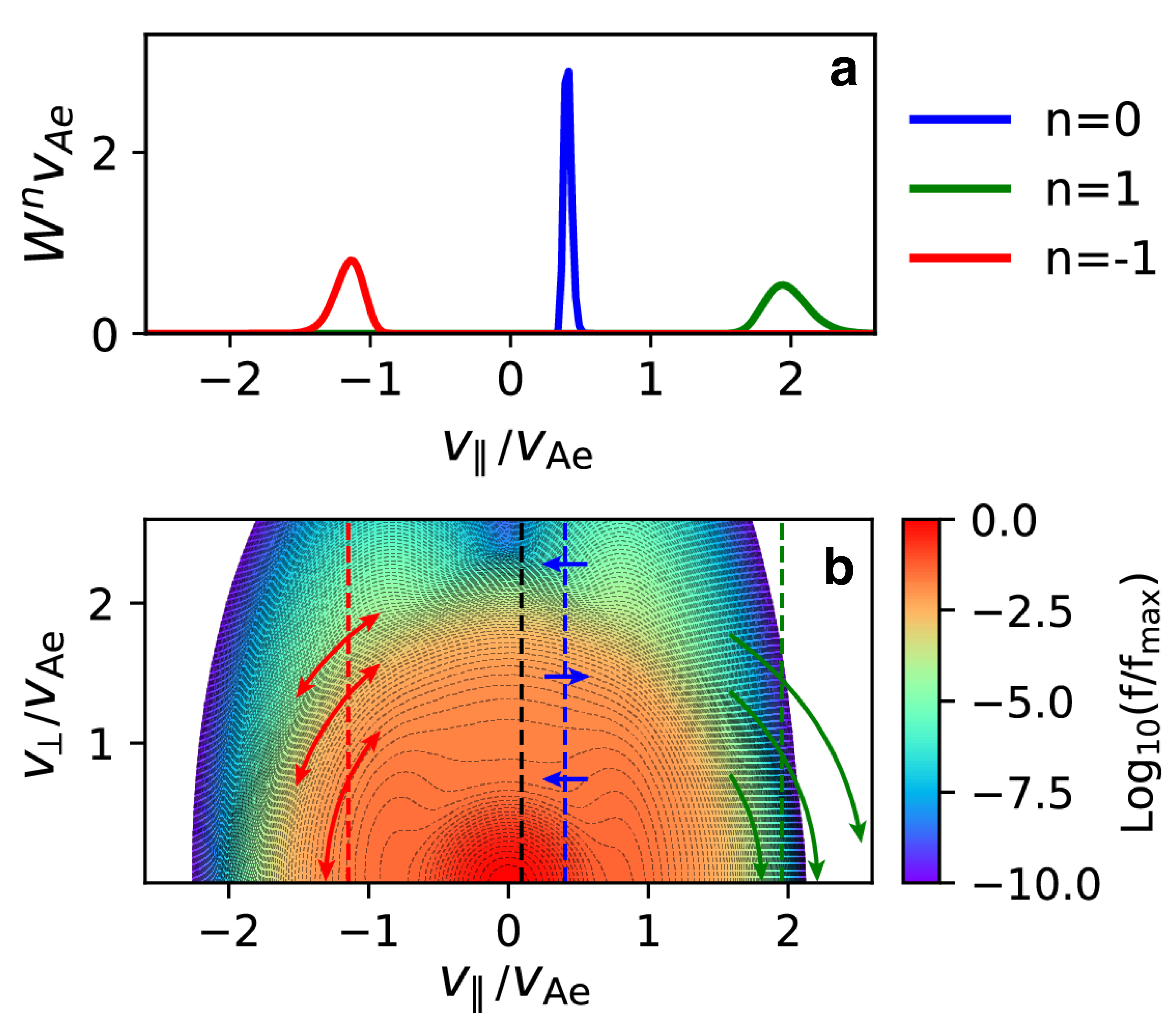}
\caption{Quasilinear diffusion in velocity space and window functions for the unstable wave energy spectrum. (a) Window functions $W^{n}$ as a function of $v_\parallel$ for $n=0$ (blue), $n=1$ (green), and $n=-1$ (red) according to Eq.~\ref{eq4}. (b) Isocontours of our electron VDF fit in velocity space. The black (blue) vertical dashed line represents the group (phase) velocity of the whistler waves. The green and red vertical dashed lines represent $v_{\parallel\mathrm{res}}^{+1}$ and $v_{\parallel\mathrm{res}}^{-1}$. The colored arrows show the local direction of the flux of diffusing particles (blue: $n=0$, green: $n=1$, and red: $n=-1$) according to  Eq.~\ref{eq6} and the local gradient of the VDF.}
\label{fig2}
\end{figure}

In Figure~\ref{fig2}a, we show the relevant $W^{n}$, which are defined to have their maximum at the resonance velocities according to Eq.~(\ref{eq5}). We determine the directions of the diffusive flux of particles according to Eq.~(\ref{eq1}) based on the local gradients of the velocity distribution function. The diffusive flux of particles in velocity space is locally tangent to {concentric elliptical/hyperbolic curves around the point $(v_{\perp}=0,v_{\parallel}=v_\mathrm{g0})$} given by \citet{Jeong2020}:
\begin{equation}
v_\perp^2+ \left[ \frac{n \Omega_\mathrm{e}}{n \Omega_\mathrm{e} - \omega_{k0} + k_\mathrm{\parallel 0}v_\mathrm{g0}} \right](v_\parallel - v_\mathrm{g0})^2= \mathrm{const.}
\label{eq6}
\end{equation}

Figure~\ref{fig2}b shows the trajectories according to Eq.~(\ref{eq6}) using colored curves with arrows for $n=0$ (blue), $n=1$ (green), and $n=-1$ (red). The arrows represent the directions of diffusive fluxes, which are always directed from larger values of $f_{\mathrm e}$ towards smaller values of $f_{\mathrm e}$. Variations in the local gradients cause the direction of the diffusive flux for $n=0$ to change at different $v_\perp$ as shown by the alternating blue arrows. 

{The direction of the diffusive flux at a specific point in velocity space is determined by the local gradient $\hat G[k_{\parallel}]f_{\mathrm e}(v_\parallel,v_\perp)$ of the VDF. It determines whether the local quasilinear diffusion contributes to wave growth or to wave damping, and can significantly vary with $v_\perp$.} If the particles lose kinetic energy during the diffusion, they contribute to the growth of the wave. If they gain energy, they contribute to its damping. Particles at different pitch angles and different energy can experience opposite effects given a specific fine structure of the VDF in velocity space. For example, around $v_\perp \approx 0.7 v_\mathrm{Ae}$ along the $n=0$ resonance, $f_{\mathrm e}$ has a negative gradient in the $v_\parallel$ direction, suggesting that the direction of the diffusive particle flux points into the negative $v_\parallel$ direction. The corresponding electrons lose kinetic energy, which thus is transferred into the wave energy. However, electrons at $v_\perp \approx 1.5 v_\mathrm{Ae}$ diffuse in the opposite direction, thus absorbing wave energy. For $n=1$,  the diffusive particle flux is clockwise at $v_{\parallel} \approx v_\mathrm{\parallel res}^{+1}$, leading to an increase in the kinetic energy of these electrons. This diffusion contributes to the damping of the whistler waves. For $n=-1$, the diffusive particle flux is counter-clockwise at $v_{\parallel} \lesssim v_\mathrm{\parallel res}^{-1}$, leading to a decrease in the kinetic energy of these electrons. This diffusion contributes to the growth of the whistler waves. Electrons at $v_{\parallel} \gtrsim v_\mathrm{\parallel res}^{-1}$ diffuse towards larger kinetic energy and thus contribute to wave damping. The net gain or loss of energy of all electrons defines whether the corresponding resonant mode undergoes growth or damping.

\subsection{Resonant Energy Transfer in Velocity Space}\label{}
Upon obtaining VDFs at different quasilinear evolution times, we evaluate the relative contributions of the three resonances to the energy conversion using the method presented by \citet{Howes2017}. We first calculate the velocity-space energy density
\begin{equation} w(v_\perp,v_\parallel,t)=\frac{1}{2}m_{\mathrm e} v^2 f_{\mathrm e}(v_\perp,v_\parallel,t) \label{eq7} 
\end{equation}
and the two-dimensional energy transfer rate 
\begin{equation} 
C(v_\perp,v_\parallel)=\frac{\partial w(v_\perp,v_\parallel,t)}{\partial t}. \label{eq8} 
\end{equation}
By performing partial integration, we obtain the one-dimensional energy transfer rates both in the parallel direction
\begin{equation} 
C_\parallel(v_\parallel)=\int_0^{\infty} 2\pi v_\perp C(v_\perp,v_\parallel) \,\mathrm d v_\perp\label{eq9} 
\end{equation}
and in the perpendicular direction
\begin{equation} C_\perp(v_\perp)=\int_\mathrm{-\infty}^{\infty} 2\pi v_\perp C(v_\perp,v_\parallel) \,\mathrm d v_\parallel. \label{eq10} 
\end{equation}
Moreover, we obtain the net energy transfer rate by a second integration over the remaining direction:
\begin{equation} C_\mathrm{net}=\int_\mathrm{-\infty}^{\infty} C_\parallel(v_\parallel) \,\mathrm d v_\parallel= \int_\mathrm{0}^{\infty} C_\perp(v_\perp) \,\mathrm d v_\perp. \label{eq11} 
\end{equation}
To compare contributions from different resonances, we define the energy transfer rates for different orders $n$ as
\begin{equation} 
C_{n}=\int_{v_\mathrm{\parallel res}^{n}-w_{n}}^{v_\mathrm{\parallel res}^{n}+w_{n}} C_\parallel(v_\parallel)\,\mathrm d v_\parallel, \label{eq12} 
\end{equation}
where $w_n$ is the half-width of the effective $v_\parallel$ range for the resonant interaction of $n$th order.

\section{Result}\label{}
\subsection{Velocity-space Signature of Energy Transfer}\label{sec31}
\begin{figure*}[!htbp]
\centering
\plotone{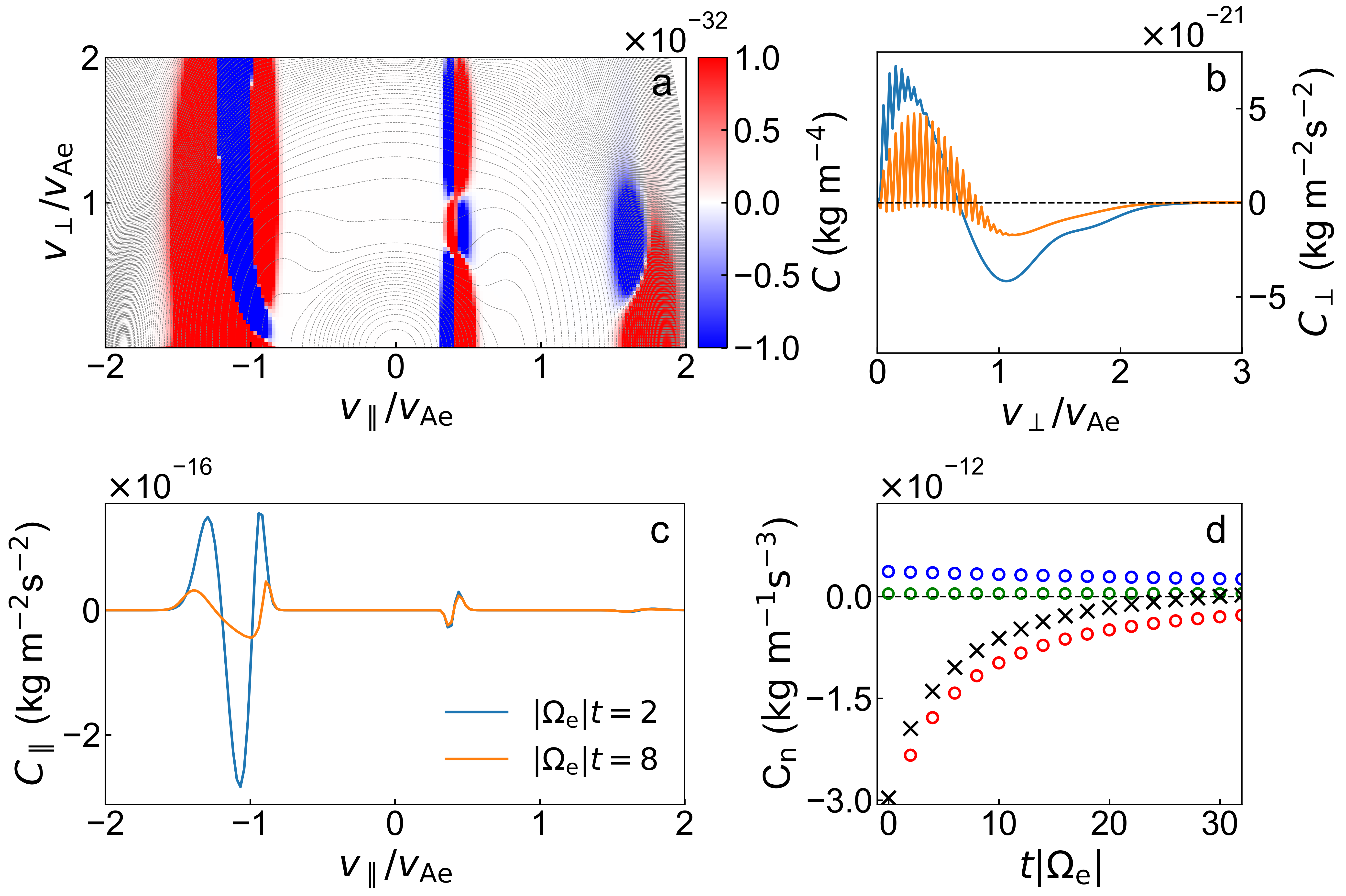}
\caption{Energy transfer rates of resonant electron--whistler interactions predicted by our numerical model. (a)  $C (v_\parallel,v_\perp)$ at $t|\Omega_{\mathrm e}|=2$ as a function of $v_\parallel$ and $v_\perp$. The background isocontours show the initial electron VDF. (b) $C_\perp (v_\perp)$ as a function of $v_\perp$. (c) $C_\parallel (v_\parallel)$ as a function of $v_\parallel$. (d) Time plots of the total energy transfer rate (black crosses) and the contributions from $n=-1$ (red circles), $n=1$ (green circles), and $n=0$ (blue circles). Blue curves in (b) and (c) represent the energy transfer rates at $t|\Omega_{\mathrm e}|=2$, and orange curves at $t|\Omega_{\mathrm e}|=8$.}
\label{fig3}
\end{figure*}

Figure \ref{fig3}a shows $C(v_\perp,v_\parallel)$ in the $(v_\perp,v_\parallel)$-plane. The isocontours represent the initial VDF, while the color represents the magnitude of $C(v_\perp,v_\parallel)$ at time $t|\Omega_\mathrm{e}|=2$. We show $C_\perp(v_\perp)$ in Figure \ref{fig3}b and $C_\parallel(v_\parallel)$ in Figure \ref{fig3}c. Different colored lines in panels (b) and (c) of Figure~\ref{fig3} represent the transfer rates  at $t|\Omega_\mathrm{e}|=2$ (blue) and  $t|\Omega_\mathrm{e}|=8$ (orange).  Figure \ref{fig3}d shows $C_\mathrm{net}$, $C_\mathrm{0}$, $C_{-1}$, and $C_\mathrm{+1}$ as functions of time.

Both one-dimensional energy conversion rates show a strong dependence on $v_\parallel$ and $v_\perp$. According to Figure \ref{fig3}a, $C(v_\perp,v_\parallel)$ shows an alternating pattern in velocity space. For $n=0$, $C(v_\perp,v_\parallel)$ shows the expected bipolar double-band signature along the $v_\parallel$ direction, which when integrated  contributes to the damping of the whistler waves. However, this bipolar signature reverses at $v_\perp \approx v_\mathrm{Ae}$ due to reversed velocity gradients of the electron VDFs along $v_\parallel$ (see also Figure~\ref{fig3}c). The reversed bipolar part of $C(v_\perp,v_\parallel)$ contributes to the growth of the whistler waves. As shown by the blue circles in Figure \ref{fig3}d, the $n=0$  resonance overall makes a damping contribution to the whistler-wave evolution. For $n=1$, $C(v_\perp,v_\parallel)$ shows a velocity-space pattern consistent with the diffusive paths shown in Figure~\ref{fig2}b. At $v_{\parallel}\approx v_{\parallel\mathrm{res}}^{+1}$, the diffusive flux is directed towards larger $v^2$, suggesting an increase in kinetic energy of the resonant electrons. While $|C_{+1}|\ll|C_0|$, the $n=1$ resonance also contributes to the damping of the whistler instability.

The $n=-1$ resonance produces a triple-band signature in $C(v_\perp,v_\parallel)$. This signature indicates that the electrons with $v_{\parallel}\approx v_{\parallel\mathrm{res}}^{-1}$ diffuse along both possible directions shown in Figure \ref{fig2}b and Figure \ref{fig3}a. Inspecting $C_\parallel(v_\parallel)$ in Figure \ref{fig3}c suggests that the electrons with $v_{\parallel}\approx v_{\parallel\mathrm{res}}^{-1}$ overall lose kinetic energy, contributing to the growth of whistler waves. The loss of kinetic energy through the $n=-1$ resonance is greater than the combined gain of kinetic energy through the $n=0$ and $n=1$ resonances. According to Figure \ref{fig3}d, the contribution of the $n=-1$ resonance is the dominant source for the growth of whistler instability.

\subsection{Quasilinear Saturation and Stabilization}\label{sec32}

During the time evolution according to Eq.~(\ref{eq1}), the magnitudes of $C_\perp(v_\perp)$ and $C_\parallel(v_\parallel)$ decrease. This result suggests that the diffusion caused by whistler waves slows down, which is a result of the decrease in the local velocity gradients of the VDFs. This secular effect is the quasilinear saturation mechanism of the whistler-wave instability under consideration. By integrating over all velocities, we obtain the net energy transfer rates at $t|\Omega_\mathrm{e}|=2$ as $C_\mathrm{-1} \approx -1.87\times 10^{-12}$\, kg\,m$^{-1}$ s$^{-3}$, $C_\mathrm{0} \approx 1.65\times 10^{-13}$\, kg\,m$^{-1}$s$^{-3}$, and $C_\mathrm{+1} \approx 2.45\times 10^{-14}$\, kg\,m$^{-1}$s$^{-3}$. We obtain a net energy transfer rate of $C_\mathrm{net} \approx 1.68\times 10^{-12}$\,kg\,m$^{-1}$s$^{-3}$. As shown in Figure~\ref{fig3}d, the net energy transfer rates gradually decrease with time, indicating quasilinear saturation of the system. Therefore, we conclude that the $n=-1$ cyclotron resonance drives the growth of the whistler waves, while the $n=0$ Landau resonance and the $n=1$ cyclotron resonance lower their growth rate.

\begin{figure*}[!htbp]
\centering
\plotone{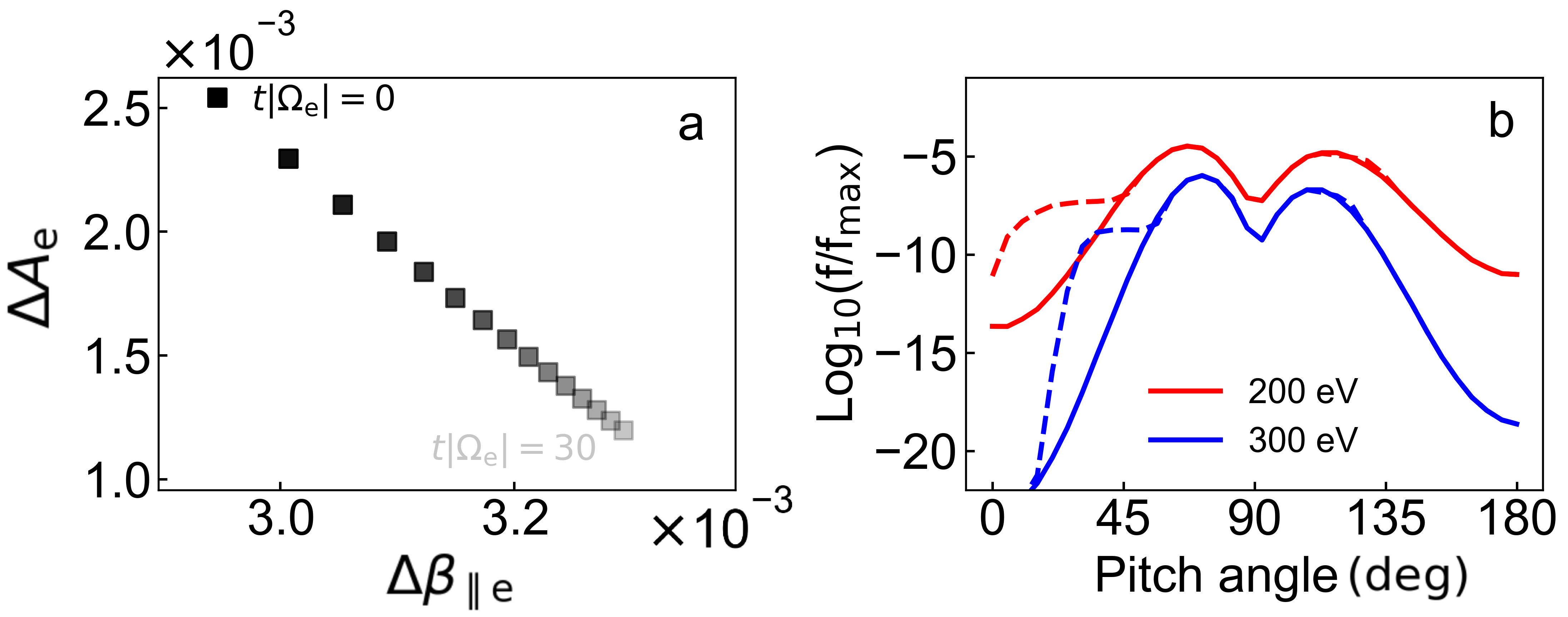}
\caption{Joint time evolution of $\Delta A_\mathrm{e}$ and $\Delta \beta_{\parallel\mathrm e}$. (a) Evolution of $\Delta A_\mathrm{e}$ and $\Delta \beta_{\parallel\mathrm e}$ at different times. (b) Pitch-angle distributions for 200~eV (red) and 300~eV (blue) electrons. Solid and dotted curves represent distributions at $t|\Omega_\mathrm{e}|=0$ and $t|\Omega_\mathrm{e}|=30$.}
\label{fig4}
\end{figure*}

Figure~\ref{fig4}a presents the time evolution of $\Delta A_\mathrm{e}$ and $\Delta \beta_{\parallel\mathrm e}$, which are the differences between $A_\mathrm{e}$ and $\beta_{\parallel\mathrm e}$ and their initial values. $A_\mathrm{e}=T_{\perp\mathrm e}/T_{\parallel\mathrm e}$ and $\beta_{\parallel\mathrm e}$ are calculated as the moments of the VDFs in our quasilinear model. From $t|\Omega_\mathrm{e}|=0$ to $t|\Omega_\mathrm{e}|=30$, $A_\mathrm{e}$ decreases and $\beta_\mathrm{e\parallel}$ increases as a result of the action of the discussed resonant wave--particle interactions. The parametric variation of $A_\mathrm{e}$ and $\beta_{\parallel\mathrm e}$ gradually decreases as the overall velocity gradients in the resonance regions decrease. The increase in $\beta_{\parallel\mathrm e}$ with time is directly related to the increase in $T_{\parallel\mathrm e}$, which is the result of the quasilinear diffusion in velocity space. The velocity-space morphology of the electron beam configuration at energies between 200 and 300~eV is noticeably weakened and evolves toward a more isotropic distribution (see Figure~\ref{fig4}b). The major contribution to the energy transfer is from regions with high phase space densities. While the contribution of the $n=1$ resonance to the total energy transfer is small, its role in reshaping the distribution in regions of velocity space with small phase space densities is still important. 

\subsection{Virtual Observations of Velocity-space Signatures}

\begin{figure}[!htbp]
\centering
\plotone{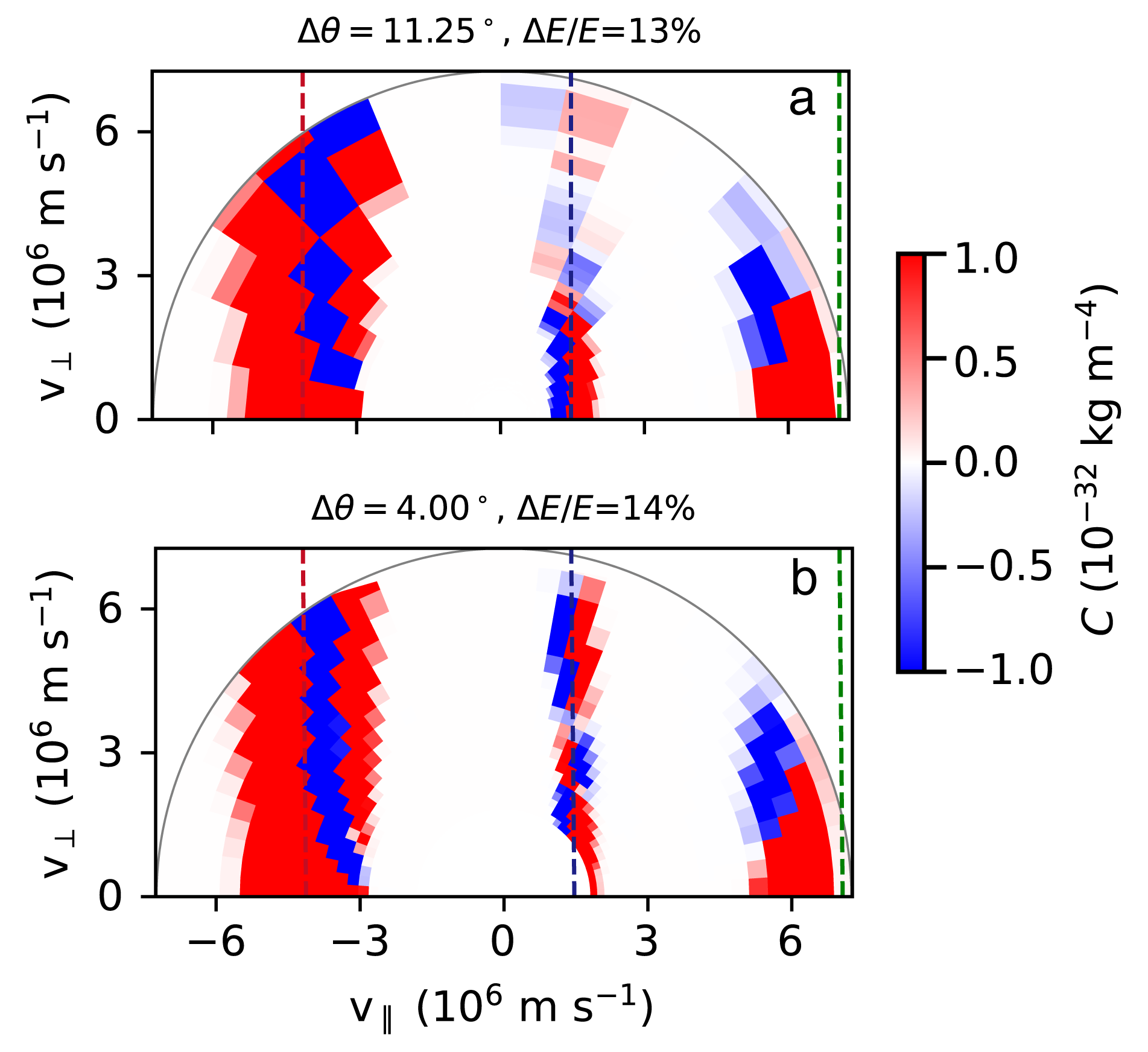}
\caption{Velocity-space signatures of energy transfer rates interpolated for different pitch-angle resolutions and energy resolutions of a particle instrument on a virtual spacecraft. We interpolate $C (v_\parallel,v_\perp)$ onto energy-angle grids with a resolution of (a) $\Delta\theta=11.25^\circ$, $\Delta E/E=13\%$, and (b) $\Delta\theta=4^\circ$, $\Delta E/E=14\%$. The blue, red, and green dashed lines represent $v_{\parallel\mathrm{res}}^0$, $v_{\parallel\mathrm{res}}^{-1}$, and $v_{\parallel\mathrm{res}}^{+1}$, respectively.}
\label{fig5}
\end{figure}

To guide future in-situ observations of similar signatures, we interpolate our result from Figure~\ref{fig3}a onto energy-angle grids with a logarithmic energy table of a virtual spacecraft. As an example, we demonstrate our results for two sets of energy-angle resolutions ($\Delta\theta=11.25^\circ$, $\Delta E/E=13\%$; and $\Delta\theta=4^\circ$, $\Delta E/E=14\%$), which are shown in Figure~\ref{fig5}. For the virtual instrument, we define the pitch-angle as $\theta$ and the energy as $E$. We assume that the background magnetic field direction is aligned with the axis when best pitch-angle resolution is achieved. The chosen energy-angle resolutions are selected according to typical values for the electrostatic analyzer onboard Solar Orbiter Electron Analyzer System (Figure~\ref{fig5}a) \citep{Owen2020} and MMS Fast Plasma Investigation (Figure~\ref{fig5}b) \citep{Pollock2016}. 

As $\Delta E/E$ and $\Delta\theta$ decrease, the signatures of the three resonances in velocity space become more pronounced and thus easier to identify. With decreasing pitch-angle resolution, the morphological features of the two-dimensional energy transfer rate become increasingly difficult to distinguish. Nevertheless, the range of wave--particle resonances are variable depending on the wave parameters.

\section{Discussion and Conclusions}\label{sec4}

Based on our MMS observations of the electron VDF in a magnetic hole, we combine linear Vlasov--Maxwell theory and a quasilinear numerical model to investigate the properties of wave--particle interactions between electrons and whistler waves. Our method reveals the velocity-space signatures of energy transfer in these resonant interactions. We quantify the relative contributions of different resonances ($n=0$, $n=-1$, and $n=1$) to the energy transfer rate in velocity space. The energy conversion rate for the $n=-1$ cyclotron resonance is the dominant contribution to the whistler wave instability, followed by the damping contributions from the $n=0$ Landau resonance and the $n=+1$ cyclotron resonance. The net energy transfer from the three resonances is $C_\mathrm{net} \approx 1.30\times 10^{-12}$\,kg\,m$^{-1}$\,s$^{-3}$. The net energy transfer from electron kinetic energy to whistler-wave energy leads to the growth of the observed electron-whistler wave instability. Furthermore, our results reveal significant dependencies of quasilinear diffusion on the local velocity-space structures of the VDFs, which gives rise to complicated patterns of energy transfer in velocity space.

Our results present complex velocity-space signatures of resonant energy transfer between unstable whistler waves and electrons. We interpolate these signatures into finite phase-space bins similar to the operating principle of recent spacecraft instrumentation such as those onboard MMS \citep{Pollock2016} and Solar Orbiter \citep{Owen2020}. The angular and energy resolution of the two instruments is sufficient for direct measurement of resonant energy transfer in velocity space. However, a significant under-sampling issue would affect the measurement if the time scale of resonant diffusion is much smaller than the time resolution of the instrument \citep{Wilson2022,Verniero2021,Horvath2022}. {We also acknowledge that the used angle/energy resolutions in Section 4.3 are simplified to some extent. The resolution of pitch angles depends on the orientation of the magnetic field direction with respect to the the instrument frame due to the conversion of instrument-frame angles (azimuth/elevation) into pitch angles. In the case of MMS, for example, the pitch-angle resolution varies between 4$^{\circ}$ and 11.25$^{\circ}$ depending on the orientation of the magnetic field.} The direct measurement of fast diffusion created by high-frequency whistler waves is still a major challenge that can potentially benefit from novel instrument concepts that focus on electron-scale wave-particle interactions. Our method, applied to high-resolution particle data from present and future missions such as Solar Orbiter, MMS, and HelioSwarm \citep{Klein2023}, is a helpful tool to investigate wave-particle interactions and their impact on VDFs both for electrons and ions. 

\acknowledgments
The Arbitrary Linear Plasma Solver (ALPS) code is publicly available at \url{https://github.com/danielver02/ALPS}, and the ALPS documentation can be found at \url{http://alps.space}. The ALPS project received support from UCL’s Advanced Research Computing Centre through the Open Source Software Sustainability Funding scheme. This research was supported by the International Space Science Institute (ISSI) in Bern, through ISSI International Team project \#563 (Ion Kinetic Instabilities in the Solar Wind in Light of Parker Solar Probe and Solar Orbiter Observations) led by L. Ofman and L.~Jian. This work is supported by NNSFC grants (42022032, 42374198, 41874203, 42188101), project of Civil Aerospace ``13th Five Year Plan" Preliminary Research in Space Science (D020301). D.V. is supported by the STFC Ernest Rutherford Fellowship ST/P003826/1 and STFC Consolidated Grants ST/S000240/1 and ST/W001004/1. H.L. is supported by the International Partnership Program of CAS (Grant No. 183311KYSB20200017) and in part by the Specialized Research Fund for State Key Laboratories of China. K.G.K. is supported by grant DE-SC0020132 and NASA ECIP 80NSSC19K0912.

\bibliography{ref}

\end{document}